\documentclass[twocolumn]{aastex7}

\usepackage{amsmath}
\usepackage{threeparttable}
\usepackage{graphicx}
\usepackage{textcomp} 
\hypersetup{allcolors=blue}

\newcommand{\LX}{$ L_{\rm X}$~}

\newcommand\iona[2]{#1$\;${\scshape{#2}}}
\newcommand{\NeV}{[Ne~{\scshape{v}}]~$\lambda3426$~}
\let\oldAA\AA
\renewcommand{\AA}{\text{\normalfont\oldAA}}

\begin{document}
\title{A
Rare Eddington-Limited, Heavily Obscured Low-Mass Active Galactic Nucleus Likely Triggered by a Galaxy Merger}

\author[0009-0003-1260-4143]{Shouyi Wang} \email[show]{sywang0302@gmail.com}
\affiliation{Department of Astronomy, School of Physics, Peking University, Beijing 100871, People's Republic of China}
\affiliation{Kavli Institute for Astronomy and Astrophysics, Peking University, Beijing 100871, People's Republic of China}
\affiliation{School of Astronomy and Space Science, Nanjing University, Nanjing, Jiangsu 210093, People’s Republic of China}
\affiliation{Key Laboratory of Modern Astronomy and Astrophysics (Nanjing University), Ministry of Education, Nanjing 210093, People’s Republic of China}

 \author[0000-0002-4436-6923]{Fan Zou} \email[show]{fanzou01@gmail.com}
\affiliation{Department of Astronomy, University of Michigan, 1085 S University, Ann Arbor, MI 48109, USA}

\author[0009-0003-4721-177X]{Chang-Hao Chen}  \email{cchen_louis@stu.pku.edu.cn}
\affiliation{Department of Astronomy, School of Physics, Peking University, Beijing 100871, People's Republic of China}
\affiliation{Kavli Institute for Astronomy and Astrophysics, Peking University, Beijing 100871, People's Republic of China}

\author[0000-0002-0167-2453]{W. N. Brandt} \email{wnbrandt@gmail.com}
\affiliation{Department of Astronomy and Astrophysics, 525 Davey Lab, The Pennsylvania State University, University Park, PA 16802, USA}
\affiliation{Institute for Gravitation and the Cosmos, The Pennsylvania State University, University Park, PA 16802, USA}
\affiliation{Department of Physics, 104 Davey Lab, The Pennsylvania State University, University Park, PA 16802, USA}

\author[0000-0001-5802-6041]{Elena Gallo}  \email{egallo@umich.edu}
\affiliation{Department of Astronomy, University of Michigan, 1085 S University, Ann Arbor, MI 48109, USA}

\author[0000-0002-9036-0063]{Bin Luo} \email{bluo@nju.edu.cn}
\affiliation{School of Astronomy and Space Science, Nanjing University, Nanjing, Jiangsu 210093, People’s Republic of China}
\affiliation{Key Laboratory of Modern Astronomy and Astrophysics (Nanjing University), Ministry of Education, Nanjing 210093, People’s Republic of China}

\author[0000-0002-7350-6913]{Xue-Bing Wu}
\email{wuxb@pku.edu.cn}
\affiliation{Department of Astronomy, School of Physics, Peking University, Beijing 100871, People's Republic of China}
\affiliation{Kavli Institute for Astronomy and Astrophysics, Peking University, Beijing 100871, People's Republic of China}

\author[0000-0002-0759-0504]{Yuming Fu}
\email{yfu@strw.leidenuniv.nl}
\affiliation{Leiden Observatory, Leiden University, Einsteinweg 55, 2333 CC Leiden, The Netherlands}
\affiliation{Kapteyn Astronomical Institute, University of Groningen, P.O. Box 800, 9700 AV Groningen, The Netherlands}

\author[0000-0002-5678-1008]{Dieu D. Nguyen}
\email{dieun@umich.edu}
\affiliation{Department of Astronomy, University of Michigan, 1085 S University, Ann Arbor, MI 48109, USA}

\author[0000-0002-1234-552X]{Shengxiu Sun} \email{sxsun@stu.pku.edu.cn}
\affiliation{Department of Astronomy, School of Physics, Peking University, Beijing 100871, People's Republic of China}
\affiliation{Kavli Institute for Astronomy and Astrophysics, Peking University, Beijing 100871, People's Republic of China}

\begin{abstract}
We report a detailed analysis of GAMA 376183, a powerful, heavily obscured active galactic nucleus (AGN) hosted by a low-mass galaxy ($M_\star \approx 10^{10}~M_{\odot}$) likely experiencing a galaxy merger. The source was initially identified due to its remarkably strong [\iona{Ne}{v}] $\lambda3426$ emission, exhibiting a rest-frame equivalent width (EW) of $\approx 48$~\AA. 
We present $\sim100$~ks Nuclear Spectroscopic Telescope Array follow-up observations, confirming its heavily obscured nature with a column density (in $\mathrm{cm^{-2}}$) of $\log N_\mathrm{H} = 23.3^{+0.4}_{-1.2}$ and an intrinsic $2$–$10$keV luminosity (in $\mathrm{erg~s^{-1}}$) of $\log L_\mathrm{X,int} = 42.92^{+0.24}_{-0.20}$. GAMA 376183 thus represents one of the few known heavily obscured AGNs in low-mass galaxies. Its estimated Eddington ratio is $\lambda_\mathrm{Edd}\approx0.8$, indicative of rapid black-hole growth. High-resolution optical images reveal a disturbed, likely merging morphology, while its multiwavelength spectral energy distribution indicates a recent starburst in its host galaxy. These pieces of evidence suggest that the ongoing merger has triggered both the heavily obscured, Eddington-limited accretion and the starburst, making GAMA 376183 a rare observed case in low-mass galaxies. Overall, this unique source demonstrates that (i) [\iona{Ne}{v}] can help identify heavily obscured low-mass AGNs, and (ii) the merger-driven coevolution framework established for massive galaxies may also extend to low-mass galaxies.

\end{abstract}

\section{Introduction}
\label{sec:intro}
Heavily obscured active galactic nuclei (AGNs), characterized by column densities of $N_\mathrm{H} \gtrsim 10^{23}~\mathrm{cm^{-2}}$, represent a critical phase in the growth of massive black holes (MBHs). Within the framework of MBH and host galaxy coevolution (e.g., \citealt{Alexander2012NewAR, Kormendy2013ARA&A}), episodes of especially vigorous mass accretion are often associated with a heavily obscured, dust-enshrouded phase, which can be triggered by galaxy mergers (e.g., \citealt{Vito2018MNRAS, Zou2020MNRAS.499.1823Z, Zou2025ApJ...990...94Z}), although mergers are not the sole mechanism. Heavily obscured AGNs constitute a substantial fraction of the overall AGN population: approximately $30\%-50\%$ of local AGNs are obscured at the Compton-thick (CT) level \citep[e.g.,][]{Boorman2025ApJ...978..118B}, and about $25\%$ of the peak power of cosmic X-ray background at $20$--$30$keV originates from CT AGNs \citep{Gilli2007A&A,Shi2013,Ananna2019ApJ}.
\par
Nearly all previous studies of heavily obscured AGNs have focused on massive galaxies with stellar masses $M_\star>10^{10}~M_\odot$ (e.g., \citealt{Lanzuisi2015A&A...578A.120L,Liu17,Ananna2019ApJ,Li19,Gilli2022A&A...666A..17G,Yan23,Mountrichas2024A&A...683A.172M,Yu24}); our understanding of these systems in lower-mass galaxies remains much more limited. Nevertheless, studying heavily obscured low-mass AGNs is important—not only as an extension of the parameter space to the low-mass regime, but also because it offers new physical insights. MBHs in low-mass galaxies may be remnants of the initial black-hole seeds, potentially preserving fossil information from the early universe (e.g., \citealt{Inayoshi2020ARA&A, Greene20}). \citet{Yang2021ApJ...921..170Y} found that most black-hole growth in the early universe likely occurs during a Compton-thick phase ($N_{\rm H}\geq1.5\times10^{24}~\rm cm^{-2}$) that is missed by current X-ray surveys. Therefore, heavily obscured low-mass AGNs at low redshifts could help reveal typical MBH growth patterns from the early universe.\par

The obscuration properties of AGNs in low-mass galaxies are poorly understood, particularly at the heavily obscured level with column densities of $N_{\rm H}\gtrsim10^{23}~{\rm cm^{-2}}$. Low-mass AGNs selected via optical spectra or variability are significantly biased against detecting obscured sources (e.g., \citealt{Reines2013ApJ,Baldassare2018ApJ,Burke22,Wasleske24}). In contrast, X-ray observations provide a powerful tool for identifying AGN activity, as X-rays can penetrate obscuring material and are much less affected by contamination from host galaxy starlight \citep{Brandt2015A&ARv}. Nevertheless, blind X-ray searches for heavily obscured low-mass AGNs remain challenging, as the discovery of a sizable sample requires both sufficient X-ray depth and coverage of a large cosmic volume. For instance, \citet{Zou2023ApJ} utilized the largest medium-depth X-ray survey to date, yet identified only a single such candidate. So far, only a few (candidate) heavily obscured low-mass AGNs have been reported \citep{Ansh2023ApJ, Zou2023ApJ, Boorman2024ApJ...975..230B,Purohit2025MNRAS,Annuar2025arXiv}.\par

Coronal lines can provide valuable additional insights and aid in selecting promising AGN candidates. These high-ionization lines arise from forbidden transitions with ionization potentials $\gtrsim 100$~eV. Coronal lines offer a unique opportunity to identify AGNs that evade conventional diagnostic methods, which are often hindered by extinction or contamination from host-galaxy star formation \citep{Satyapal2021ApJ}. Among these, \NeV is one of the most promising and well-studied lines (e.g., \citealt{8Gilli2010,11Mignoli2013A&A...556A..29M,6Cleri2023ApJ...953...10C,7Cleri2023ApJ...948..112C}). Its production requires ionizing photon energies as high as $97.11$~eV, which is seven times greater than the hydrogen ionization energy and well above the lower bound of the ``very high" ionization zone defined by \citet{Berg2021ApJ...922..170B}. Photons at such high energies can even be considered ultra-soft X-rays. 
\citet{Reiss2025ApJ...989...88R} analyzed a large sample of low-redshift, ultrahard X-ray-selected, narrow-line AGNs detected by Swift-BAT  and found that their \NeV emission lines were robustly detected in approximately 43\% (146/341) of the sources. Notably, the detection rate remains high ($\approx60\%$) even in heavily obscured sources with column densities reaching $\log N_H / \text{cm}^{-2} \approx 24$.
\citet{Peca2025ApJ...990....3P} investigated a sample of highly luminous and obscured AGNs, reporting a \NeV detection rate of $85\%$ (17/20) among sources with available optical spectroscopy. 
Besides, a correlation between intrinsic $2$--$10$~keV X-ray luminosity ($L_\mathrm{X}$) and [\iona{Ne}{v}]~$\lambda3426$ luminosity ($L_{\rm [Ne\; \text{\scshape v}]}$) has been found by \citet{8Gilli2010} and \citet{Reiss2025ApJ...989...88R}.
Therefore, [\iona{Ne}{v}] may help reveal the missing, heavily obscured low-mass AGN population as well, especially given that the incidence of coronal lines is higher in low-mass galaxies compared to massive ones \citep{Reefe2023ApJ...946L..38R}.\par

However, identification of low-mass AGNs via [\iona{Ne}{v}] has been precluded in the past due to limited spectral coverage. In particular, spectroscopically identified low-mass AGNs in the Sloan Digital Sky Survey (SDSS) generally reside at very low redshifts \citep[$z \lesssim 0.06$; e.g.,][]{Reines2013ApJ}; for these objects, SDSS spectra do not cover [\iona{Ne}{v}]. At higher redshifts, survey sensitivity becomes a significant challenge. The Galaxy and Mass Assembly survey \citep[GAMA;][]{Driver2011MNRAS.413..971D,Liske2015MNRAS.452.2087L,Bellstedt2020MNRAS,GAMADR42022MNRAS.513..439D}, which is two magnitudes deeper than SDSS, enables spectroscopic selection of low-mass AGNs beyond the local universe, with [\iona{Ne}{v}] redshifted into the GAMA spectral range. \citet{Salehirad2022ApJ} systematically searched the entire GAMA database and identified 388 low-mass AGNs. Among the three sources with [\iona{Ne}{v}] detections reported in their study, GAMA 376183 ($z=0.209$) has the largest [\iona{Ne}{v}] equivalent width (EW $\approx$ 52 \AA),  while the remaining two sources show lower values, with EW$ < 10$ \AA. We therefore select this source for follow-up analyses.\par

We present detailed analyses of this newly discovered candidate. We observed GAMA 376183 with the Nuclear Spectroscopic Telescope Array \citep[NuSTAR;][]{Harrison2013ApJ} for approximately 100~ks, confirming its heavy obscuration. Multiwavelength analyses further suggest that this AGN is likely merger-driven. The detailed methodology and results are presented below. Section~\ref{sec: data} describes the data analyses and source properties of GAMA 376183. Section~\ref{sec: discussion} discusses the physical implications. Section~\ref{sec: summary} summarizes the main findings of this work. Throughout this paper, we use J2000 coordinates and adopt a $\Lambda$CDM cosmology with $H_0=70~\mathrm{km~s^{-1}~Mpc~^{-1}}$, $\Omega_\Lambda = 0.7$, and $\Omega_M = 0.3$. Luminosities are always quoted in $\mathrm{erg~s^{-1}}$, and $N_\mathrm{H}$ is given in units of $\mathrm{cm}^{-2}$.

\section{Multi-wavelength Data and Properties of GAMA 376183}
\label{sec: data}

\subsection{Optical Spectroscopy}
\label{sec: optspec}
\begin{figure*}
  \centering
  \includegraphics[width=\textwidth,height=2.5in]{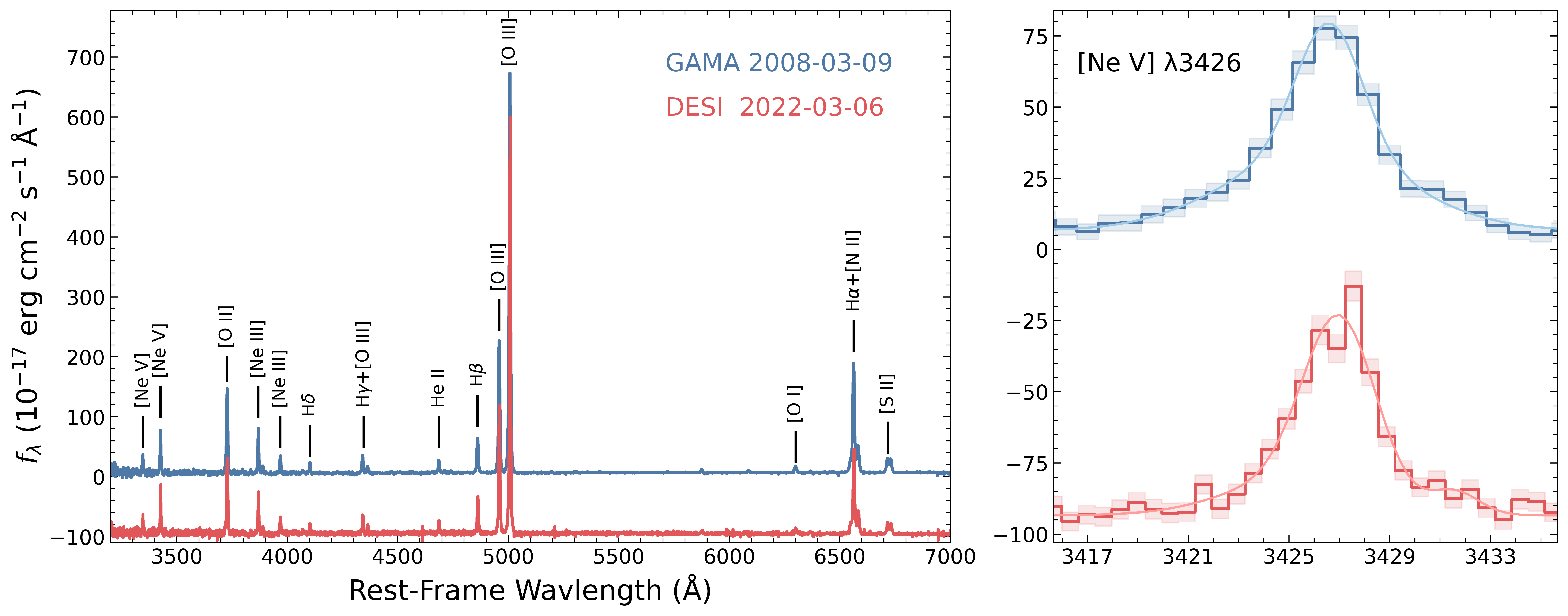}
  \caption{Optical spectra of GAMA 376183. Apparent emission lines are marked explicitly. The DESI spectrum is shifted downward  by $1\times 10^{-15}~\rm erg~cm^{-2}~s^{-1}~\AA^{-1}$ for display purposes. The right panel presents a zoom-in of the [\iona{Ne}{v}] emission line. The observed spectrum is displayed with a step function, the shaded region indicates its uncertainties, and the solid curve represents the best-fit model, which consists of a power-law continuum plus two Gaussian components. \label{fig:opspec}}
\end{figure*}

GAMA 376183 was first identified by the GAMA survey and was later reobserved by the Dark Energy Spectroscopic Instrument (DESI; \citealt{DESIDR1}) survey in March 2008 and March 2022, respectively. The GAMA and DESI spectra are presented in Figure~\ref{fig:opspec}. Both spectra are corrected for Galactic extinction using the extinction law from \citet{extinclaw2019ApJ...886..108F}, with $R_{V}=3.1$ and a Galactic $E(B-V)=0.037$ \citep{Schlegel1998ApJ...500..525S}.\par

\begin{figure*}
  \centering
  \includegraphics[width=\textwidth,height=2.5in]{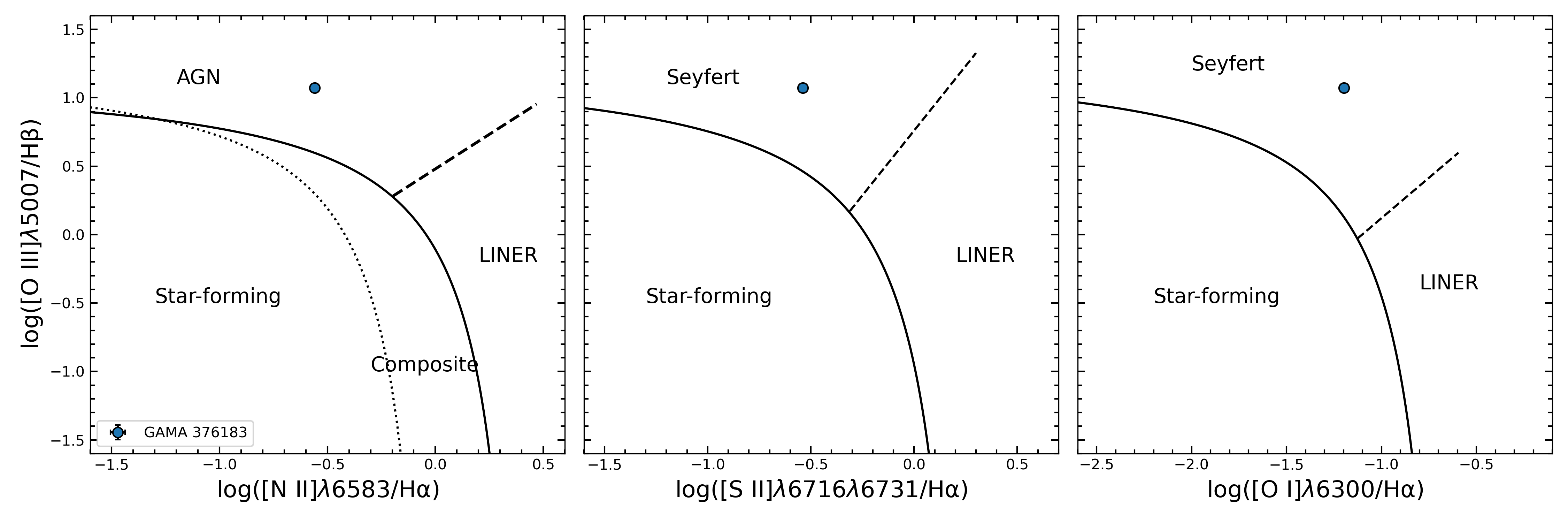}
  \caption{The BPT diagram with demarcations determined by \citet{Kewley2006MNRAS.372..961K}. Our source is clearly identified as an AGN in these diagrams.\label{fig:BPT}}
\end{figure*}

GAMA 376183 exhibits no apparent variability between the two epochs, and its AGN nature is confirmed by the BPT diagram \citep[see Figure~\ref{fig:BPT};][]{Baldwin1981PASP...93....5B,Veilleux1987ApJS...63..295V}.
Although the SED fitting with \texttt{CIGALE} suggests that the galaxy may have experienced a recent starburst episode (see Section~\ref{sec:SED}), the emission-line diagnostics indicate that the ionization is dominated by an AGN. The source lies in the AGN region of the BPT diagram, where it is difficult for pure starburst models to reproduce the observed line ratios \citep{Kewley2001ApJ...556..121K}.
In addition, the detection of the high-ionization [Ne V] emission line further supports the presence of an AGN. We fit [\iona{Ne}{v}] within the 3380--3480~\AA\ spectral window using a power-law continuum and two Gaussian components with \texttt{QSOFITMORE} \citep{QSOFIT2018ascl, Shen2019ApJS, QSOFITMORE}. The GAMA (DESI) spectrum yields a [\iona{Ne}{v}] luminosity of $\log L_{\rm [Ne\; \text{\scshape v}]} = 41.67^{+0.12}_{-0.16}$ ($41.60^{+0.14}_{-0.20}$), and an [\iona{Ne}{v}] EW of $52.0 \pm 16.1$ ($44.2 \pm 16.4$)~\AA. Since these measurements are broadly consistent between the two spectra, we adopt the averaged EW value for subsequent analysis.

\subsection{Image Decomposition}
\label{sec: imgdecomp}
GAMA 376183 was observed as part of the Hyper Suprime-Cam Subaru Strategic Program (HSC-SSP) Wide survey \citep{Aihara2022_hsc-ssp_dr3} in five bands ($grizy$), covering rest-frame wavelengths between $0.5$ and $1~\mu{\rm m}$. The high spatial resolution of the HSC images enables detailed modeling of galaxy substructures, while the broad optical spectral energy distribution (SED) coverage allows accurate derivation of the stellar mass for each component.\par
For this analysis, we utilize \texttt{GalfitS} (R. Li \& L. C. Ho, in preparation), an open-source Python package designed for simultaneous multiband image analysis, which self-consistently incorporates both spatial and SED information. \texttt{GalfitS} facilitates definition of parametric profiles for each model component, such as a S\'ersic profile for extended galaxy structures and a point source for nuclear emission. A physically motivated SED model is assigned to each component, generating a spatially resolved SED map that is convolved with the filter transmission curves to produce multiband image models. With user-provided cutout images, error maps, and point spread functions (PSFs) in each band, \texttt{GalfitS} calculates the likelihood function via $\chi^2$ statistics and simultaneously optimizes the morphological and SED parameters within a Bayesian framework. The code has already been employed by \citet{Chen_C2025b, Chen_C2025a} and \citet{Li_R2025} for measuring host-galaxy properties of high-redshift AGNs.

We obtain $13.44\arcsec \times 13.44\arcsec$ cutout images and corresponding error maps from the HSC-SSP\footnote{\url{https://hsc-release.mtk.nao.ac.jp/doc/index.php/data-access__pdr3/}} Data Release 3. The point spread functions (PSFs) were generated using PSF Picker,\footnote{\url{https://hsc-release.mtk.nao.ac.jp/psf/pdr3/}} which constructs PSF models from stars near the target location with the \texttt{PSFEx} software \citep{psfex}. We began with single-band fitting of the HSC $g$-band image to assess the necessary profiles for describing the morphology and to estimate initial parameters. Two sets of models were run: one employing a single S\'ersic profile, and the other using two S\'ersic profiles with their center positions tied, decomposing the target into possible disk and bulge components. In both cases, an additional point-source component was included to account for potential nuclear emission.

\begin{figure*}
  \centering
  \includegraphics[width=\textwidth]{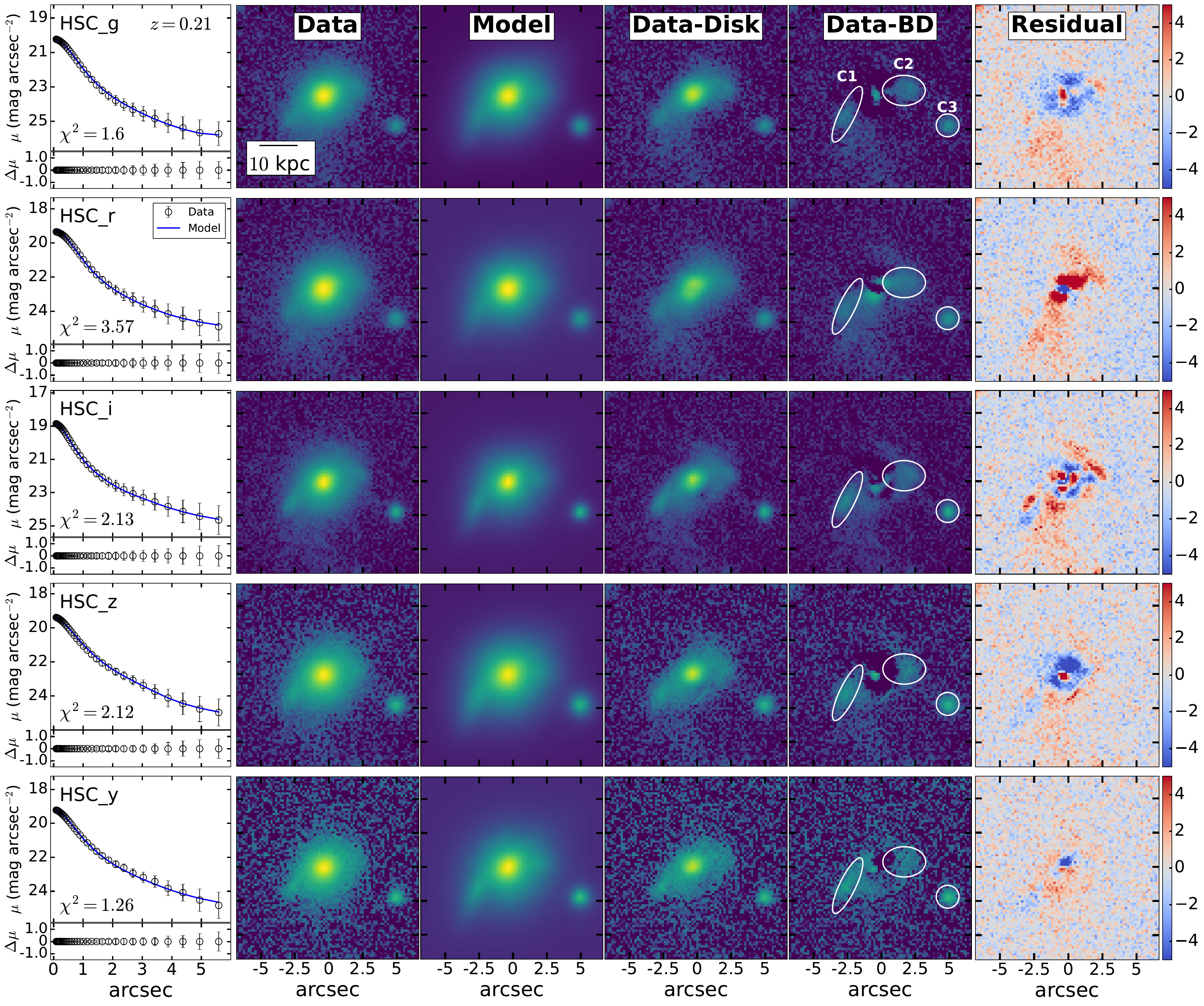}
  \caption{Simultaneous multiband image fitting results for GAMA 376183. Each row corresponds to one HSC band. In the leftmost column, the upper panel displays the radial surface brightness profile of the target (open circles with error bars) compared to the best-fit model (blue solid line); the $\chi^2$ value for each band is reported in the lower left corner. The lower panel shows the residuals between the data and the best-fit model. The imaging columns, from left to right, present the original data, best-fit model, data minus the disk component, data minus both disk and bulge components, and the residuals normalized by the errors (data-model/error). The positions of the subcomponents C1, C2, and C3 are highlighted by white circles in the fourth imaging column. \label{fig:image_decomposition}}
\end{figure*}

The two-component model yielded a significant decrease in the fitting Bayesian Information Criterion (BIC) by $\approx 900$, indicating the necessity of bulge/disk decomposition. After subtracting the best-fit profiles from the data, several substructures became apparent, labeled as C1, C2, and C3 in Figure~\ref{fig:image_decomposition}. C1 and C2 appear to be associated with GAMA 376183, with their elongated and twisted morphologies suggesting possible interactions or merger events. C3, located farther from GAMA 376183, can be well described by an unresolved point source. C3 has an $i$-band magnitude of 22.56 (see later), and \citet{Bosch18} showed that a point-like source in HSC images at this magnitude is most likely a foreground star because galaxies with $i$-band magnitudes brighter than $\approx24$ should generally be resolvable by HSC. We further fit C1 and C2 using additional S\'ersic profiles and modeled C3 with a point-source component.

With the profiles and initial parameters determined from the single-band fitting, we proceed with the formal analysis by simultaneously fitting the morphology and SED models across all five HSC bands. In this formal run, we adopt four S\'ersic profiles (bulge, disk, C1, and C2) and two point sources (nucleus and C3), initializing each parameter with its best-fit value from the single-band fitting.

Each of the four S\'ersic profiles is assigned a galaxy SED model, composed of theoretical stellar templates from the binary population and spectral synthesis (BPASS V2.3; \citealt{bpass}) code, assuming a stellar initial mass function (IMF) from \citet{kroupa_imf}, and nebular emission templates computed with CLOUDY \citep{Cloudy}. The star formation history (SFH) is modeled with seven non-parametric bins: the most recent bin spans 100 Myr, while the remainder are equally divided in logarithmic space, assuming star formation began 200 Myr after the Big Bang. Stellar masses for each component are permitted to vary between $10^6$ and $10^{12}\ M_{\odot}$, with initial values based on a $g$-band mass-to-light ratio of 1. Dust attenuation $A_{\rm V}$ is allowed in the range $0$–$5$, metallicity $Z$ between $0.01$ and $2\ Z_{\odot}$, and the ionization parameter $U$ for nebular emission is fixed at ${\rm log}\ U=-2$.

The fluxes associated with the two point sources (nucleus and C3) are $1$–$2$ magnitudes fainter in all five bands compared to the bulge and disk components. This negligible contribution from the nuclear point source, representing the central AGN, is further supported in Section~\ref{sec:SED}, which demonstrates that the AGN SED is heavily obscured, resulting in minimal optical emission. As our primary objective is to derive the stellar masses of all galaxy components, we do not assign any SED model to the point sources; thus, their multiband fluxes are calculated directly from the fitted point source models.\par

The fitting results from the formal run are presented in Figure~\ref{fig:image_decomposition}. The bulge in GAMA 376183 has an effective radius of $R_{\rm e} = 0.26~{\rm kpc}$, a large S\'ersic index ($n = 4.05$), and a stellar mass of $\log(M_\mathrm{bulge}/M_{\odot}) = 9.40 \pm 0.02$. In contrast, the disk is more extended, with $R_{\rm e} = 0.48~{\rm kpc}$, relatively flat ($n = 2.01$), and has a stellar mass of $\log(M_\mathrm{disk}/M_{\odot}) = 10.03 \pm 0.21$, albeit with a larger uncertainty due to its lower surface brightness. Combined, the bulge and disk yield a total stellar mass of $\log M_\star = 10.12 \pm 0.17$, statistically consistent with the value of $\log M_\star = 9.98$ reported by \citet{Salehirad2022ApJ}. The stellar masses of C1 and C2 are subdominant relative to the bulge and disk, with a combined mass ratio of $M_{\star,\rm C1+C2}/M_{\star,\rm B+D} = 0.10$. If C1 and C2 are indeed merging components, this value would represent the lower limit of the initial merger mass ratio because, since there is only one galaxy nucleus, the likely merger should be in the late stage where most of the stellar mass have coalesced. The magnitudes of all of the components are summarized in Table~\ref{tab:compomag}.\par

\begin{deluxetable*}{c c c c c c c c}
\setlength{\tabcolsep}{8pt} 
\tablecaption{Component magnitudes with $1\sigma$ uncertainties.\label{tab:compomag}}
\tablehead{
\colhead{Band} &
\colhead{Bulge} &
\colhead{Disk} &
\colhead{AGN} &
\colhead{C1} &
\colhead{C2} &
\colhead{C3} &
\colhead{$\lambda_\mathrm{eff}$ [\AA]}
}
\startdata
$g$ &
$20.49_{-0.02}^{+0.01}$ &
$20.74_{-0.42}^{+1.52}$ &
$21.44_{-0.20}^{+0.25}$ &
$23.84_{-0.02}^{+0.17}$ &
$23.46_{-0.18}^{+0.06}$ &
$24.29_{-0.02}^{+0.02}$ &
4816.12 \\
$r$ &
$19.94_{-0.04}^{+0.01}$ &
$19.76_{-1.91}^{+0.41}$ &
$19.77_{-0.01}^{+0.01}$ &
$22.85_{-0.05}^{+0.19}$ &
$22.77_{-0.15}^{+0.22}$ &
$23.10_{-0.01}^{+0.01}$ &
6234.11 \\
$i$ &
$19.83_{-0.02}^{+0.01}$ &
$19.36_{-1.20}^{+0.66}$ &
$20.28_{-0.04}^{+0.90}$ &
$22.26_{-0.06}^{+0.26}$ &
$22.47_{-0.04}^{+0.27}$ &
$22.56_{-0.01}^{+0.01}$ &
7740.58 \\
$z$ &
$19.64_{-0.02}^{+0.02}$ &
$19.13_{-2.07}^{+0.63}$ &
$21.51_{-0.58}^{+0.71}$ &
$21.96_{-0.03}^{+0.40}$ &
$22.28_{-0.05}^{+0.06}$ &
$22.38_{-0.01}^{+0.01}$ &
9125.20 \\
$y$ &
$19.61_{-0.02}^{+0.03}$ &
$19.13_{-1.56}^{+0.49}$ &
$21.33_{-0.15}^{+0.39}$ &
$21.81_{-0.06}^{+0.19}$ &
$22.22_{-0.05}^{+0.17}$ &
$22.16_{-0.02}^{+0.02}$ &
9779.93 \\
\enddata
\end{deluxetable*}

We have also tested to measure photometric redshifts of C1 and C2 with \texttt{LePhare} \citep{Arnouts99, Ilbert06}. The adopted galaxy templates are the same as in \citet{Ilbert09}, and we add a nominal 5\% systematic photometric uncertainty to each band. The derived photometric redshifts are loosely constrained to be $0.53_{-0.44}^{+0.19}$ and $0.30_{-0.29}^{+0.42}$ for C1 and C2, respectively. The larger uncertainties essentially only return upper limits for photometric redshifts. Such limited constraints are not surprising because, if C1 and C2 are also at our source's spectroscopic redshift of 0.209, their Balmer breaks would not be covered by the HSC bandpass, causing their $grizy$ SEDs to lack strong features for reliable photometric redshift measurements. Nevertheless, these measurements are at least not in conflict with the implicit assumption that C1 and C2 belong to the overall system with same redshifts.

\subsection{UV-to-IR SED Fitting}\label{sec:SED}
In this section, we fit the full ultraviolet (UV) to infrared (IR) SED to analyze both the galaxy and AGN components. The available photometric bands generally do not resolve the detailed morphological structures shown in Figure~\ref{fig:image_decomposition}; therefore, we use only cumulative fluxes. This approach is justified, as Section~\ref{sec: imgdecomp} demonstrates that the total light is dominated by the main galaxy. Compared to the optical data alone in Section~\ref{sec: imgdecomp}, the broader-wavelength SED offers complementary information on the most recent star formation (primarily traced by UV) and AGN emission (primarily traced by IR), as discussed in detail below.\par

For SED fitting, we employ \texttt{CIGALE} v2025.0 \citep{CIGALE2019}. Although \texttt{CIGALE} allows the inclusion of X-ray data \citep{Yang2020MNRAS.491..740Y}, we do not utilize the X-ray module, as it relies on an empirical relation between X-ray and UV luminosities calibrated for AGNs in massive galaxies. This relation may not be applicable to low-mass AGNs \citep[e.g.,][]{Dong2012ApJ...755..167D} and could yield inaccurate modeling results. All photometric measurements are taken from the input catalog of GAMA \citep{Bellstedt2020MNRAS}, including the Galaxy Evolution Explorer \citep[GALEX;][]{Martin2005ApJ...619L...1M}, the Sloan Digital Sky Survey \citep[SDSS;][]{York2000AJ....120.1579Y}, Visible and Infrared Survey Telescope for Astronomy \citep[VISTA;][]{Sutherland2015A&A...575A..25S}, Wide-field Infrared Survey Explorer \citep[WISE;][]{Wright2010AJ....140.1868W}, and the Photodetector Array Camera and Spectrometer (PACS) on the Herschel Space Observatory \citep{Pilbratt2010A&A...518L...1P,Poglitsch2010A&A...518L...2P}. The measurements are summarized in Table~\ref{tab:allprop} and corrected for Galactic extinction.

\begin{table}
\begin{footnotesize}
\renewcommand{\thetable}{\arabic{table}}
\centering
\caption{Observed properties of GAMA 376183	}
\label{tab:allprop}
\begin{threeparttable}
\begin{tabular}{lr}
\tablewidth{0pt}
\hline
\hline
GAMA 376183		&								\\
\hline
R.A. (GAMA) [J2000, deg] 			& 	132.3135					\\
Decl. (GAMA) [J2000, deg]			& 	1.4885 				\\
Redshift 		&	0.209			\\
{GALEX} FUV  [$\mu$Jy]	&	57.38 	$\pm$ 	5.74  \\
{GALEX} NUV  [$\mu$Jy]	&	35.95 	$\pm$ 	3.63			\\
SDSS $u$-band [$\mu$Jy]		&	34.11	$\pm$	3.84			\\
SDSS $g$-band [$\mu$Jy]		&	64.87	$\pm$	6.52			\\
SDSS $r$-band [$\mu$Jy]		&	178.01	$\pm$	17.81			\\
SDSS $i$-band [$\mu$Jy]		&	159.86	$\pm$	16.24			\\
VISTA $Z$-band [mJy]	&	0.16 $\pm$ 0.02				\\
VISTA $Y$-band [mJy]	&	0.18 $\pm$ 0.02				\\
VISTA $J$-band [mJy]	&	0.21 $\pm$ 0.02				\\
VISTA $H$-band [mJy]	&	0.26 $\pm$ 0.03				\\
VISTA $ Ks$-band [mJy]	&	0.32 $\pm$ 0.03				\\
{WISE} 3.4 $\micron$ [mJy]	&	0.26 $\pm$ 0.03				\\
{WISE} 4.6 $\micron$ [mJy]	&	0.36 $\pm$ 0.04				\\
{WISE} 12  $\micron$ [mJy]	&	2.05 $\pm$ 0.21				\\
{WISE} 22  $\micron$ [mJy]	&	10.50 $\pm$ 1.06			\\
PACS 100  $\micron$ [mJy]	&	35.71 $\pm$ 10.73			\\
PACS 100  $\micron$ [mJy]	&	23.00 $\pm$ 8.92			\\
\hline
Image fitting with \texttt{GalfitS} (Section ~\ref{sec: imgdecomp})\\
\hline
$\log M_{\star,\mathrm{B}}$ & 9.40\\
$\log M_{\star,\mathrm{D}}$ & 10.03\\
$M_{\star,\rm C1+C2}/M_{\star,\rm B+D}$ & 0.10\\
\hline
SED fitting with \texttt{CIGALE} (Section ~\ref{sec:SED}) &								\\
\hline
$\log M_\star$ [$M_{\sun}$]\tnote{a}								&	10\\
SFR [$M_{\sun}$ yr$^{-1}$]						&	$8.36 \pm 3.07 $	\\
AGN fraction &$0.89 \pm 0.07$ \\
Viewing angle [degree] & $89 \pm 1$ \\
$\log L_{6\mu m}$ [erg s$^{-1}$]		&	$43.80_{-0.06}^{+0.04}$\\
$\log L_{\rm bol}$ [erg s$^{-1}$]	 & $44.83^{+0.04}_{-0.05}$ \\

\hline
X-ray spectral analysis	  (Section ~\ref{sec: xray}) &								\\
\hline
Effective $\Gamma$ & $0.92^{+0.5}_{-0.5}$ \\
$\log N_{\rm H}$ [cm$^{-2}$]							&	$23.3^{+0.4}_{-1.2}$		\\
$\log L_{\rm X, obs}$ (2-10 keV)  [erg s$^{-1}$]			& 	$42.42^{+0.17}_{-0.20}$		\\ 
$\log L_{\rm X, int}$ (2-10 keV)  [erg s$^{-1}$] & $42.92^{+0.24}_{-0.20}$ \\

\hline
BH properties	  (Section ~\ref{sec:mergepropp}) &								\\
\hline
$\log M_{\rm BH}$ [$M_{\sun}$]						& 	$6.83\pm0.30$			\\ 
$\lambda_{\rm Edd}$								&	$0.81\pm0.33$ \\

\hline
\end{tabular}
    \begin{tablenotes}
      \item[a] The stellar masses in \citet{Salehirad2022ApJ}, our optical resolved SED fitting in Section~\ref{sec: imgdecomp}, and UV-to-FIR SED fitting in Section~\ref{sec:SED} are $10^{9.98}~M_{\sun}$, $10^{10.12}~M_\odot$, and $10^{10.25}~M_{\sun}$, respectively. Thus, we adopt a nominal $M_\star=10^{10}~M_{\sun}$.
    \end{tablenotes}
\end{threeparttable}
\end{footnotesize}
\end{table}

We employ the AGN model \texttt{skirtor2016} \citep{Stalevski2016MNRAS.458.2288S}, the dust emission model \texttt{dale2014} \citep{Dale2014ApJ...784...83D}, the attenuation model \texttt{dustatt\_modified\_starburst} \citep{Calzetti2000ApJ...533..682C}, the nebular emission model \texttt{nebular}, the SFH model \texttt{sfhdelayed} (i.e., the classical delayed SFH) or \texttt{sfhdelayedbq} (a delayed SFH with additional recent bursting or quenching events; see \citealt{Ciesla2016A&A...585A..43C}), and the stellar template model \texttt{bc03} \citep{Bruzual2003MNRAS.344.1000B}. Note that the IMF in \texttt{GalfitS} is fixed to \citet{kroupa_imf} and cannot be changed, while \texttt{CIGALE} (in our setup) does not offer a Kroupa IMF option. Because the IMF models in \citet{kroupa_imf} and \citet{Chabrier2003ApJ...586L.133C} yield very similar stellar mass-to-light ratios, we therefore adopt the \citet{Chabrier2003ApJ...586L.133C} IMF in \texttt{CIGALE}.  The specific model parameters adopted are summarized in Table~\ref{tab:cigale}.

\begin{deluxetable*}{llll}

\tablecaption{\texttt{CIGALE} parameter settings for GAMA 376183 \label{tab:cigale}}
\tablehead{
\colhead{Module} & \colhead{Parameter} & \colhead{Name in \texttt{CIGALE}} & \colhead{Possible values}
}
\startdata
Delayed SFH & Stellar \textit{e-}folding time & tau\_main & 0.1, 0.3, 0.5, 0.7, 0.9, \\
&&& 1, 3, 5, 7, 9, 10 Gyr \\
& Stellar age & age\_main & 0.1, 0.3, 0.5, 0.7, 0.9, \\
&&& 1, 3, 5, 7, 9, 10 Gyr \\
(or)&&&\\
Delayed SFH with burst & Stellar \textit{e-}folding time & tau\_main & 0.1, 0.3, 0.5, 0.7, \\
&&& 1, 3, 5, 7, 10 Gyr \\
& Stellar age & age\_main & 1, 3, 5, 7 Gyr \\
&Age of the BQ episode &age\_bq & 5, 10, 50, 100, 200, 300\\
&&& 400, 500, 600, 700, 800 Myr\\
& Ratio of the SFR after/before age\_bq& r\_sfr& 0, 1, 5, 10, 30, 50\\
&&&100, 500, 1000\\
\hline
Simple stellar population & Initial mass function & imf & \citet{Chabrier2003ApJ...586L.133C} \\
& Metallicity & metallicity & 0.0001, 0.0004, 0.004, \\
&&& 0.008, 0.02, 0.05 \\
\hline
Nebular & ----- & ----- & ----- \\
\hline
Dust attenuation & $E(B-V)_\mathrm{line}$ & E\_BV\_lines & 0, 0.1, 0.2, 0.3, 0.4, 0.5, \\
&&& 0.6, 0.7, 0.8, 0.9, 1, 1.2, 1.5 \\
& $E(B-V)_\mathrm{line}/E(B-V)_\mathrm{continuum}$ & E\_BV\_factor & 1 \\
\hline
Dust emission & Alpha slope & alpha & 1.0, 1.5, 2.0, 2.5, 3.0 \\
\hline
AGN & Viewing angle & i & $0^\circ, 10^\circ, 30^\circ, 50^\circ, 70^\circ, 90^\circ$ \\
& Disk spectrum & disk\_type & \citet{Schartmann2005} \\
& Modification of the optical power-law index & delta & $-0.27$ \\
& AGN fraction & fracAGN & 0, 0.05, 0.1, 0.2, 0.3, 0.4, \\
&&& 0.5, 0.6, 0.7, 0.8, 0.9, 0.99 \\
& $E(B-V)$ of the polar extinction & EBV & 0, 0.05, 0.1, 0.2, 0.3, 0.4, 0.5 \\
\enddata
\tablecomments{Each module name is listed only in the first relevant row, and all multi-line values are preserved as in the original table. Unlisted parameters are set to the default values. }
\end{deluxetable*}

\par
\begin{figure}
  \centering
  \includegraphics[width= 3.3in,height=2.3in]{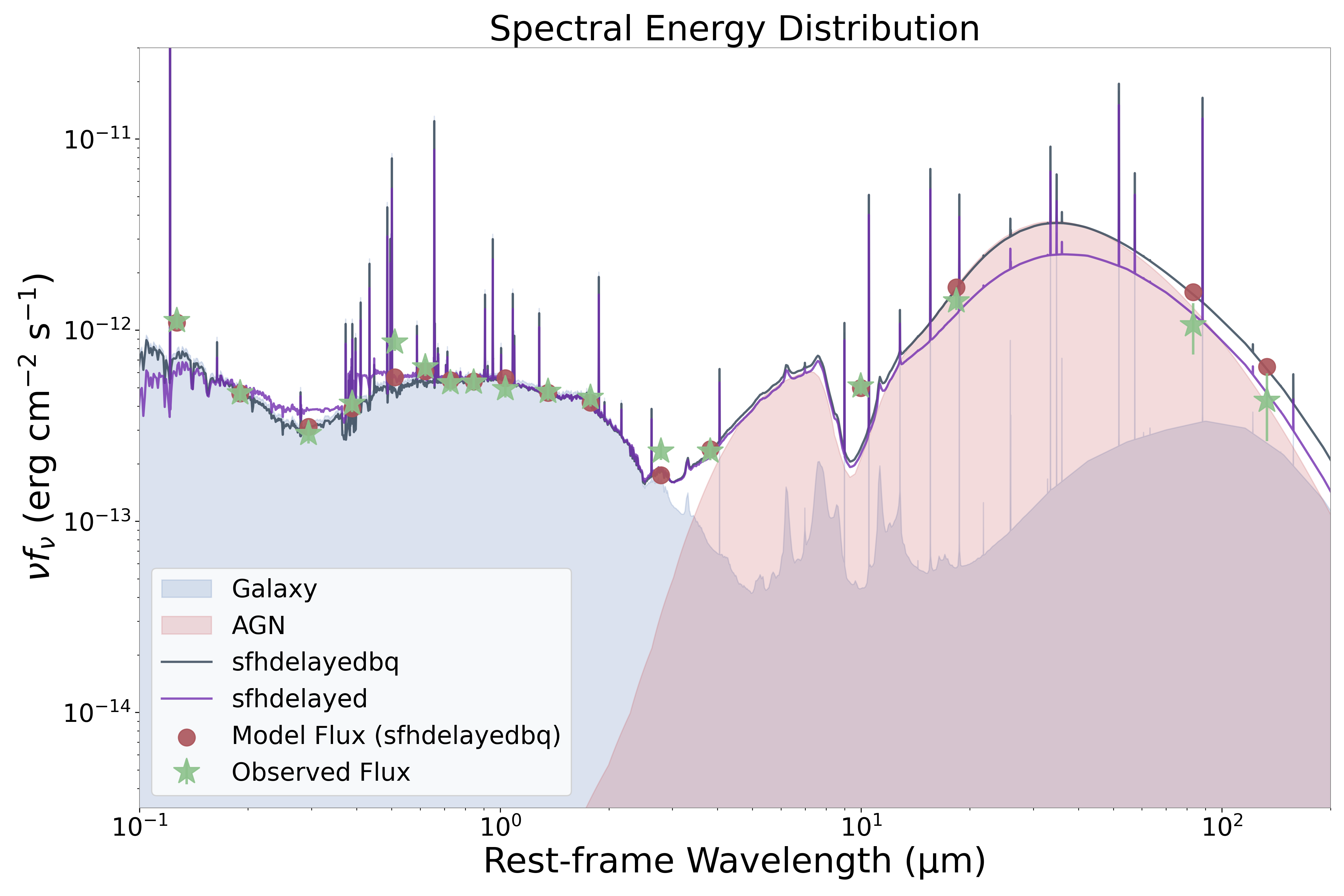}
  \caption{Results of SED fitting for GAMA 376183, with observed photometric data points (green stars) spanning from the UV to the far-IR. The SED is decomposed into galaxy (blue) and AGN (red) components. The black and purple line show the total SED for \texttt{sfhdelayedbq} and \texttt{sfhdelayed} SFH models, respectively. \label{fig:sed}}
\end{figure}

The SED fitted with \texttt{sfhdelayed} is plotted as the purple line in Figure~\ref{fig:sed}; the fit yields a $\chi_{\rm reduced}^2 = 2.7$, where the degrees of freedom\footnote{\texttt{CIGALE} defines the degrees of freedom as the number of photometric data points minus one, where ``minus one'' stands for the fact that only one parameter, the model normalization, is free at each grid point.} is 16. However, the current model underpredicts the FUV flux relative to the observations, with the observed UV SED much bluer than the fitted model, indicating that the instantaneous SFR at $\lesssim10-100$ Myr timescales is high—evidence for a recently triggered starburst. Thus, we replace the SFH module \texttt{sfhdelayed} with \texttt{sfhdelayedbq} \citep{Ciesla2016A&A...585A..43C} to allow for a recent burst episode that can reproduce an increase in the UV output. Such a model is also physically motivated for a merger-driven system, which may induce enhancements in star formation \citep{DiA&A...468...61D,CoxMNRAS.384..386C,Willett2015MNRAS.449..820W}.  The adoption of the \texttt{sfhdelayedbq} model enhances the fit quality, yielding a $\chi_{\rm reduced}^2$ of 1.8, and successfully reproduces the observed  UV flux. Especially, the combined $\chi^2$ for the FUV, NUV, and $u$ bands significantly improves from 13.6 to 0.8 after using \texttt{sfhdelayedbq}. $\chi_{\rm reduced}^2=1.8$, as a nominal fitting quality indicator, indicates that the overall fitting is acceptable. As a comparison, the \texttt{CIGALE} fitting of Seyfert galaxies in \citet{RamosPadilla22}, covering the same UV to FIR range as our SED, has a typical range of $0.3\lesssim\chi_{\rm reduced}^2\lesssim3$. Additionally, one third of the UV-to-FIR SED fits of galaxies brighter than GAMA~376183 in \citet{Zou2022ApJS..262...15Z} also have $\chi_{\rm reduced}^2>1.8$. Therefore, our fitted $\chi_{\rm reduced}^2$ is typical and hence acceptable for our purposes. 

Therefore, we will adopt the new SFH hereafter. Figure~\ref{fig:sed} also shows that the AGN continuum is significantly suppressed in the UV to near-infrared bands, supporting the scenario of heavy obscuration.\par

The galaxy and AGN properties estimated from the \texttt{CIGALE} SED fitting are summarized in Table~\ref{tab:allprop}. The stellar mass is $\log M_\star = 10.25_{-0.12}^{+0.09}$, consistent with the optical SED-based value ($\log M_\star = 10.12 \pm 0.17$) in Section~\ref{sec: imgdecomp} and the value ($\log M_\star = 9.98$) from \citet{Salehirad2022ApJ}, considering the typical systematic uncertainty of 0.2~dex in SED-based $M_\star$ estimates (e.g., \citealt{Pacifici2023ApJ...944..141P}). Accordingly, we adopt a nominal value of $M_\star = 10^{10}~M_\odot$ hereafter. The derived star formation rate (SFR) is $8.36 \pm 3.07~M_{\odot}~\mathrm{yr^{-1}}$, placing GAMA 376183 on the star-forming main sequence according to \citet{Whitaker2012ApJ}. As a comparison, the ProSpectAGNv02 table provided by GAMA returns SFR = $3.59~M_{\odot}~\mathrm{yr^{-1}}$, and the SFR derived using our \texttt{sfhdelayed} model is $6.34 \pm 0.65~M_{\odot}~\mathrm{yr^{-1}}$. Given the typical systematic uncertainty of SED-based SFR measurements is 0.3~dex \citep{Pacifici2023ApJ...944..141P}, we regard our \texttt{sfhdelayed} model-based SFR to be consistent with the independent SFR measurement by GAMA. However, according to the \texttt{sfhdelayedbq} SFH fitting, a recent starburst occurred with a fitted age of the burst episode of $96\pm210~\rm Myr$ and an SFR increasing factor of $40\pm98$ at the burst. It is not surprising that the burst-parameter uncertainties are large because it is notably difficult to accurately constrain these high-order SFH quantities with broadband SEDs (e.g., see \citealt{Suess22} for a detailed discussion). Nevertheless, the main evidence for the recent starburst originates from the UV part of the SED, and the improvement of the FUV, NUV, and $u$-band $\chi^2$ from 13.6 to 0.8 after adding the starburst component is indeed statistically significant. The fractional AGN contribution to the total dust emission is $0.89\pm 0.07$, and the AGN viewing angle is $89^{\circ} \pm 1^{\circ}$, consistent with the fact that GAMA 376183 is a type 2 AGN. The SED decomposition yields an AGN rest-frame $6~\mu$m luminosity of $\log L_{6\mu\mathrm{m}} = 43.80_{-0.06}^{+0.04}$ and an AGN bolometric luminosity of $\log L_{\rm bol} = 44.83^{+0.04}_{-0.05}$.

\subsection{X-Ray Analyses}
\label{sec: xray}
GAMA 376183 was observed by NuSTAR twice in December 2024, with exposure times of 61.6~ks and 47.7~ks for the first and second observations, respectively (see Table~\ref{tab:detection} for details). Data are processed using the standard NuSTAR pipeline \citep[HEASoft v6.34 with NuSTARDAS v2.1.4a][]{HEAsoft2014ascl.soft08004N}, with \texttt{NUPIPELINE} employed to generate cleaned and calibrated event files. To address background enhancement caused by the South Atlantic Anomaly (SAA), we apply an additional SAA filter (\texttt{SAAMODE = OPTIMIZED}, \texttt{TENTABLE = YES}, and \texttt{SAACALC = 1}) during \texttt{NUPIPELINE} processing. The second observation is flagged for elevated solar activity, which increases detector background by up to a factor of two to three during flare episodes. Solar flares are filtered out using a custom Python tool\footnote{\url{https://github.com/NuSTAR/nustar-gen-utils/blob/main/notebooks/GTI_filter_solar_flare.ipynb}} developed by the NuSTAR team. The resulting cleaned exposure times are 59.8~ks and 46.2~ks for the first and second observations, respectively. For both the FPMA and FPMB detectors, we produce 3--24~keV (full band) NuSTAR images using the \texttt{NUPRODUCTS} task.\par
We define the source region as a $40''$-radius circle centered on the optical position of GAMA 376183, and the background region as three source-free $60''$-radius circles located on the same detector chip. Within the source region, GAMA 376183 has 278 and 224 total counts in the 3--24~keV band for the first and second observations, respectively. 
To assess the significance of the source detection, we compute the binomial no-source probability $P_\mathrm{B}$, defined as
\begin{equation} \label{equ1}
P_{\mathrm{B}} = \sum_{X=S}^{N} \frac{N!}{X!(N-X)!} p^{X} (1-p)^{N-X},
\end{equation}
where $S$ is the total number of counts in the source region, $B$ is the total number of counts in the background region, $N = S + B$, and $p = 1/(1 + \mathrm{BACKSCAL})$, with BACKSCAL being the ratio of the background and source region areas. $P_\mathrm{B}$ represents the likelihood that the observed source counts are due to background fluctuations, assuming that no real source is present at the specified location. We consider the source detected if $P_\mathrm{B} < 0.01$ (corresponding to a $\geq 2.6~\sigma$ detection).

Table~\ref{tab:detection} summarizes the results of the source-detection algorithm for the 3–8 keV (soft), 8–24 keV (hard), and 3–24 keV (full) bands. Net counts are also reported in Table~\ref{tab:detection}. For bands and observations where the source is detected, $1\sigma$ uncertainties of the net counts are calculated following Section 1.7.3 of \citet{Lyons1991pgda.book}; otherwise, we report 90\% confidence-level upper limits following \citet{Kraft1991ApJ}. GAMA 376183 is detected in the hard and full bands during the first observation; although not formally detected in the soft band, a marginal signal may be present with $P_\mathrm{B} < 0.05$. We have also verified that the source net counts keep increasing as we progressively include higher-energy bands, indicating that the source signal should be present up to higher energies. In the second observation, no detection is achieved in any band due to elevated overall background levels from solar flares, even after we filtered out the strongest flares. Especially, the background-region count rate increases by $\approx$ 30\% in the second exposure, which suppresses the source detection significance. Therefore, the subsequent spectral analysis will be based solely on the first observation.

\begin{deluxetable*}{llccccclccc}
\tablecaption{NuSTAR Photometric Properties \label{tab:detection}}
\tablehead{
\colhead{Obsid} & \colhead{Observation} & \colhead{Exposure Time} &
\multicolumn{3}{c}{Net Source Counts} && \multicolumn{3}{c}{$\rm P_{\rm B}$} \\
\colhead{} & \colhead{Start Date} & \colhead{(ks)} &
\colhead{3--8 keV} & \colhead{8--24 keV} & \colhead{3--24 keV} && 
\colhead{3--8 keV} & \colhead{8--24 keV} & \colhead{3--24 keV} \\
\colhead{(1)} & \colhead{(2)} & \colhead{(3)} &
\colhead{(4)} & \colhead{(5)} & \colhead{(6)} &&
\colhead{(7)} & \colhead{(8)} & \colhead{(9)}
}
\startdata
61001007002 & 2024--12--03 & 59.8 & $<42.0$ & $27.9^{+16.9}_{-15.9}$ & $50.5^{+23.7}_{-22.4}$ && 0.031 & 0.009 & 0.001 \\
61001007004 & 2024--12--07 & 46.2 & $<21.4$ & $<13.3$ & $<22.8$ && 0.463 & 0.771 & 0.648 \\
\enddata

\tablecomments{
\textit{Notes.} Columns (1)--(2): NuSTAR observation ID and observation date. Column (3):  Cleaned exposure time. Columns (4)--(6): Net counts in the soft band (3--8 keV), hard band (8--24 keV) and full band (3--24 keV). An upper limit at a 90\% confidence level is given if GAMA 376183 is not detected in that band. Columns (7)--(9): Binominal no-source probability in the soft, hard, and full bands.
}
\end{deluxetable*}

We further extract the source and background spectra using the \texttt{NUPRODUCTS} task and merge the FPMA and FPMB spectra and response files using the HEASoft tool \texttt{ADDSPEC}. The 3--24 keV NuSTAR spectrum is fitted using \texttt{sherpa} \citep{Siemiginowska24} with the $W$-statistic adopted to account for the limited photon counts. The Galactic neutral hydrogen column density is fixed at $3.23 \times 10^{20} \rm~cm^{-2}$ (\citealt{HI4PICollaboration}). To start, we adopt a simple power-law model modified by Galactic absorption to describe the spectrum. The model is \texttt{phabs*clumin*powerlaw}, where \texttt{powerlaw} modeled the X-ray continuum; \texttt{phabs} describes the Galactic absorption, and \texttt{clumin} calculates the observed luminosity. The power-law photon index derived from this modeling, which serves as a basic proxy of the overall spectral shape, is $\Gamma=0.92^{+0.5}_{-0.5}$ ($W$ = 175.9 for 195 degrees of freedom). The corresponding observed $2-10$~keV luminosity is $\log L_{\rm X,obs}=42.42^{+0.17}_{-0.20}$.

The hard spectral slope above 3~keV suggests strong absorption, warranting an assessment of whether the direct line-of-sight absorption column toward the nuclear X-ray source is Compton thick. We further added an intrinsic absorption component, and the model is \texttt{phabs*zphabs*zpowerlw}, where \texttt{zphabs} accounted for the intrinsic absorption. We fixed $\Gamma$ to 2, and the best fit ${\rm log}N_{\rm H} = 23.3^{+0.4}_{-1.0}$ with intrinsic $2$--$10$~keV luminosity $\log L_{\rm X,int} = 42.85^{+0.19}_{-0.18}$ ($W = 178.8$ for 195 degrees of freedom).

We also tried the BORUS model in \citet{Borus+2018} for the absorption and reflection with the following model configuration: \texttt{phabs*(borus+zphabs*cabs*cutoffpl)}. Here, \texttt{phabs} accounts for the foreground Galactic absorption, while \texttt{zphabs*cabs} models line-of-sight absorption and Compton scattering losses at the redshift of the source. The column density parameter, $N_\mathrm{H}$, is linked across \texttt{borus}, \texttt{zphabs}, and \texttt{cabs}. The \texttt{cutoffpl} component represents the intrinsic continuum, and its parameters are linked to the corresponding ones in the \texttt{borus} model. We fixed $\Gamma$ to 2, the relative iron abundance to 1, and the continuum cutoff energy to 300 keV for the fitting.

\begin{figure}
  \centering
  \includegraphics[width= 3.3in,height=2.3in]{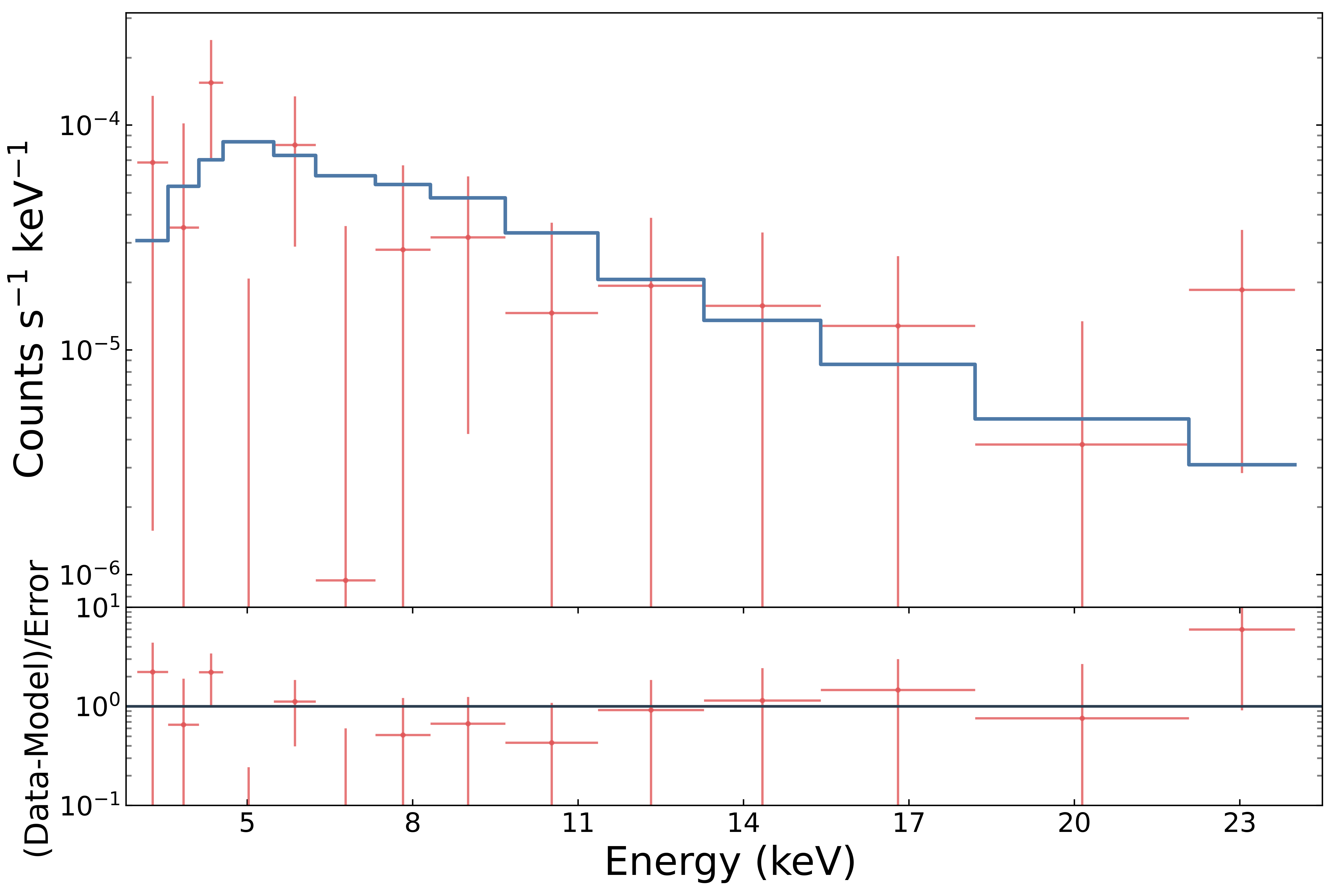}
  \caption{Background-subtracted NuSTAR spectrum, overlaid with the best-fit BORUS model. The bottom panel displays the fitting residuals. The data are grouped for display purposes only. We note that the background-subtracted count rate remains positive at high energies (>10 keV), indicating that the source still has signals in the hard X-rays.\label{fig:xspec}}
\end{figure}

The best-fit BORUS model for the spectrum, shown in Figure~\ref{fig:xspec}, yields $\log N_{\rm H} = 23.3^{+0.4}_{-1.2}$ and an intrinsic $2$--$10$~keV luminosity of $\log L_{\rm X,int} = 42.92^{+0.24}_{-0.20}$ ($W = 179.1$ for 195 degrees of freedom). The measured $N_\mathrm{H}$ supports the idea that GAMA 376183 is a heavily obscured AGN. The uncertainty of $\log L_{\rm X,int}$ is smaller than that of $\log N_\mathrm{H}$ because, at $\log N_\mathrm{H}\approx23.3$, the absorption of $2$--$10$~keV X-rays remains only moderate ($\lesssim0.6$~dex). The fitting results are summarized in Table~\ref{tab:Xfit}.

To compare the fitting quality of these models, we rely on the Bayesian information criterion (BIC) because our models are not nested, and hence the classical $F$ test cannot be applied. Since the numbers of free parameters of these models are all two, their BIC differences $\Delta\mathrm{BIC}$ is thus $\Delta W\le3.2$, suggesting that three spectral models provide comparable fit statistics and are indistinguishable from a pure statistical view. However, BORUS offers a more physically motivated treatment of obscuration geometry and reflection, so we adopt this model for our physical interpretation. Nevertheless, the BORUS model is effectively similar to the absorbed power-law model because the reflection component becomes important only when $N_\mathrm{H}\gg10^{24}~\mathrm{cm^{-2}}$.

\begin{deluxetable*}{lccc}
\tablecaption{\hbox{Best-fit results with different models}} \label{tab:Xfit}
\tablewidth{0pt} 

\tablehead{
\colhead{Parameters} & \colhead{Model 1} & \colhead{Model 2} & \colhead{Model 3} \\
\colhead{} & \colhead{\texttt{phabs*powerlaw}} & \colhead{\texttt{phabs*zphabs*powerlaw}} & \colhead{\texttt{phabs*(borus+zphabs*cabs*cutoffpl)}}
}
\startdata
$\Gamma$ & $0.92_{-0.5}^{+0.5}$ & 2 (fixed) & 2 (fixed) \\
$\log N_\mathrm{H}$ & -- & $23.3_{-1.0}^{+0.4}$ & $23.3_{-1.2}^{+0.4}$ \\
Normalization$^a$ & $1.8^{+3.9}_{-1.3}\times10^{-6}$ & $2.4^{+1.3}_{-0.8}\times10^{-5}$ & $2.5^{+1.8}_{-0.9}\times10^{-5}$ \\
\textit{W}/\textit{d.o.f.} & 175.9/195 & 178.8/195 & 179.1/195 \\
$\log L_{\rm X, obs}$ (erg s$^{-1}$) & $42.42^{+0.17}_{-0.20}$ & -- & -- \\
$\log L_{\rm X, int}$ (erg s$^{-1}$) & -- & $42.85^{+0.19}_{-0.18}$ & $42.92^{+0.24}_{-0.20}$ \\
$\log F_{\rm X, obs}$ (erg s$^{-1}$ cm$^{-2}$)& $-14.32^{+0.17}_{-0.20}$& -- & -- \\
$\log F_{\rm X, int}$ (erg s$^{-1}$ cm$^{-2}$)& -- & $-13.21^{+0.19}_{-0.18}$ & $-13.19^{+0.24}_{-0.20}$ \\
\enddata

\tablecomments{$^a$ Power-law normalization at observed-frame 1~keV in units of $\rm keV^{-1}~cm^{-2}~s^{-1}$. The errors correspond to 68\% confidence interval. For \texttt{phabs*(borus+zphabs*cabs*cutoffpl)} model, the column density parameter, $N_\mathrm{H}$, is linked across \texttt{borus}, \texttt{zphabs}, and \texttt{cabs}. The relative iron abundance is fixed to 1, and the continuum cutoff energy is fixed to 300 keV.}
\end{deluxetable*}

\section{Discussion}
\label{sec: discussion}
\subsection{Comparison between Different Inferred Luminosities}
\label{sec: comp_lumin}
\begin{figure}
  \centering
  \includegraphics[width= 3.3in,height=2.3in]{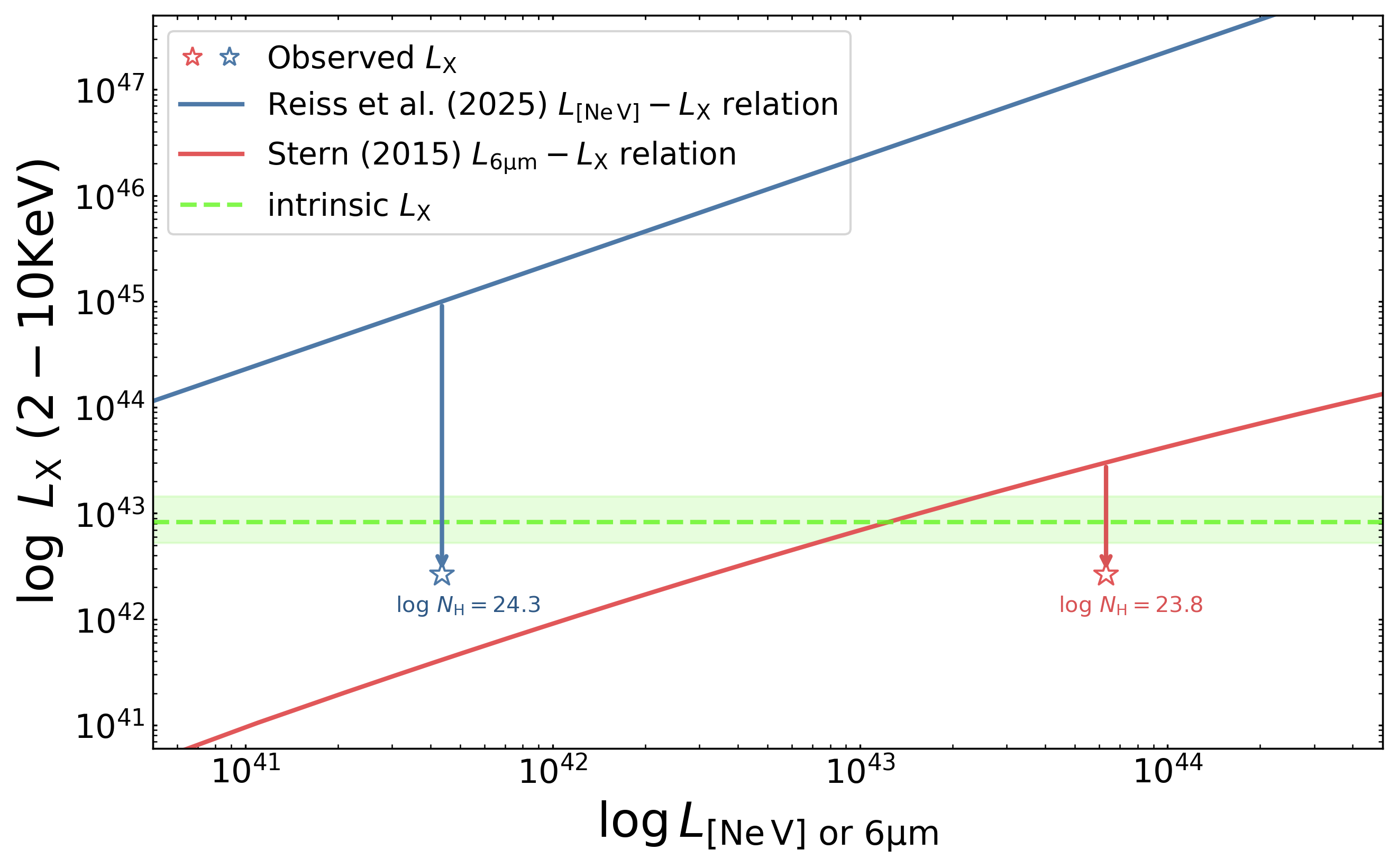}
  \caption{Comparison between the scaling relations between  the intrinsic 2–10 keV luminosity and various AGN  luminosity indicators, including  \LX  vs. $L_{\rm [Ne\; \text{\scshape v}]}$ (blue; \citealt{Reiss2025ApJ...989...88R}) and \LX vs.  $L_{6\mu m}$ (red; \citealt{Stern2015ApJ...807..129S}). The intrinsic X-ray luminosity derived based on the \texttt{BORUS} model is shown as the horizontal dashed green line with $1\sigma$ uncertainties marked as the shaded region. The vertical arrows show the deviation between the observed X-ray luminosity and intrinsic X-ray luminosity based on the scaling relations. We also estimate the  column density needed to cause this deviation using the \texttt{phabs*cabs*powerlaw} model.     \label{fig:LXrelation}} 
\end{figure}

In this subsection, we investigate whether the observed and intrinsic $2$–$10$~keV luminosities of GAMA 376183 are consistent with those inferred from its $L_{6\mu \mathrm{m}}$ and $L_{\rm [Ne\; \text{\scshape v}]}$. Such comparisons provide insight into the reliability of alternative diagnostics for identifying heavily obscured or CT AGNs when combined with X-ray observations. The comparisons are illustrated in Figure~\ref{fig:LXrelation}. The solid blue line represents the relation between the intrinsic $L_\mathrm{X}$ and $L_{\rm [Ne\; \text{\scshape v}]}$ from \citet{Reiss2025ApJ...989...88R}, $L_\mathrm{X}/L_{\rm [Ne\; \text{\scshape v}]} = 10^{3.36}$, which has a typical scatter of 0.47~dex. Similarly, the solid red line shows the relation between $L_\mathrm{X}$ and $L_{6\mu\mathrm{m}}$ from \citet{Stern2015ApJ...807..129S}, ${\rm log}~L_{\rm {X}}=40.981 + 1.024(\log L_{6\mu\mathrm{m}}-41)-0.047(\log L_{6\mu\mathrm{m}}-41)^2$, with a typical scatter of 0.4~dex.
Figure~\ref{fig:LXrelation} shows that our observed X-ray luminosity $L_\mathrm{X,obs}$ lies significantly below both relations. Based on the \texttt{phabs*cabs*powerlaw} model, this suppression corresponds to $\log N_\mathrm{H} = 24.3_{-0.1}^{+0.1}$ and $23.8^{+0.2}_{-0.3}$, where the $N_\mathrm{H}$ uncertainties are propagated from the luminosity uncertainties., if $L_\mathrm{X,int}$ were exactly on the $L_{\rm [Ne\; \text{\scshape v}]}$–$L_\mathrm{X}$ and $L_{6\mu\mathrm{m}}$–$L_\mathrm{X}$ relations, respectively, which consistently indicate that GAMA 376183 is subject to heavy obscuration.

\par

When we compare our intrinsic X-ray luminosity $L_\mathrm{X,int}$ from X-ray spectral fitting with these relations, the value is somewhat lower than expected.
For the $L_{6\mu\mathrm{m}}$–$L_\mathrm{X}$ relation, the difference is only at a $1\sigma$ level, given the typical intrinsic scatter of 0.4~dex. However, our ratio $L_\mathrm{X,int}/L_{\rm [Ne\; \text{\scshape v}]}$ is only 20, far below the typical value of $10^{3.36}$ reported by \citet{Reiss2025ApJ...989...88R}; this corresponds to a deviation greater than $3\sigma$, considering the relation's intrinsic scatter of 0.47~dex. 
This discrepancy, combined with the  large [\iona{Ne}{v}] EW, likely reflects a combination of geometric and physical effects. In this Type 2 AGN, the dusty torus obscures the central continuum, while leaving the  [\iona{Ne}{v}] emission from the narrow-line region largely unaffected, naturally elevating the EW and suppressing the $L_{\rm X}/L_{\rm [Ne\; \text{\scshape v}]}$ ratio. Furthermore, a lack of significant host-scale dust attenuation in GAMA 376183 could allow the [Ne V] line to escape efficiently. These factors, alongside an enhancement of coronal lines in low-mass galaxies \citep[e.g.,][]{Cann2018ApJ...861..142C, Reefe2023ApJ...946L..38R}, potentially explain the observed [\iona{Ne}{v}] strength.
\par
To further assess whether the intrinsic X-ray luminosity predicted from the $6~\mu\mathrm{m}$ and [\iona{Ne}{v}] indicators is physically feasible, we perform an additional test using the \texttt{phabs*(borus+zphabs*cabs*cutoffpl)} model. Specifically, we compute the normalization corresponding to the intrinsic luminosity inferred from each relation and refit the X-ray spectrum with this normalization fixed, leaving $N_\mathrm{H}$ as the only free parameter. When adopting the $6~\mu\mathrm{m}$–based intrinsic luminosity, the best‐fit yields $\log N_\mathrm{H} = 24.1_{-0.1}^{+0.1}$ and a \textit{W}/\textit{d.o.f.}$=183.8/196$. In contrast, the [Ne \textsc{v}]–based luminosity pushes the fit to the model upper limit of $N_\mathrm{H} = 10^{24.8}~\mathrm{cm^{-2}}$, with a much poorer \textit{W}/\textit{d.o.f.}$=410.5/196$.
These results suggest that the intrinsic luminosity implied by the $6~\mu\mathrm{m}$ relation is broadly consistent with the X-ray data, even though the data quality is insufficient to constrain such high obscuration directly. Physically, GAMA 376183 remains compatible with the $L_{6\mu\mathrm{m}}-L_\mathrm{X}$ relation and is likely to host a CT AGN.

\subsection{\NeV as a Tracer of AGN activity in  Heavily obscured, Low-mass systems}
\label{sec:Nevew}

In addition to GAMA 376183, \citet{Zou2023ApJ} identified another candidate heavily obscured AGN, XMM-LSS 02399, in a low-mass galaxy from the XMM-Spitzer Extragalactic Representative Volume Survey \citep{Chen2018MNRAS.478.2132C, Ni2021ApJS..256...21N}, the largest medium-depth X-ray survey to date. Although XMM-LSS 02399 is only a candidate due to less reliable $M_\star$ measurements (see their Appendix B), it also displays a strong [\iona{Ne}{v}] line with an EW of 35~\AA. 

Originating in the narrow-line region, [\iona{Ne}{v}] emission persists even in heavily obscured AGNs where X-ray emissions are suppressed, providing a robust proxy for identifying these hidden systems \citep{11Mignoli2013A&A...556A..29M,Hickox2018ARA&A..56..625H,Vergani2018A&A...620A.193V,Negus2023ApJ...945..127N}.
Both GAMA 376183 and XMM-LSS 02399 suggest that the AGN diagnostic power of [\iona{Ne}{v}] up to the heavily obscured regime  also extends to low-mass systems.

As discussed in Section~\ref{sec:intro}, this method has limited effectiveness in previous spectroscopic surveys of low-mass candidates due to constraints in sensitivity and spectral coverage. However, with the current or upcoming much deeper surveys such as DESI \citep{DESIDR1}, PFS \citep{Tamura2016SPIE.9908E..1MT}, and 4MOST \citep{deJong2019Msngr.175....3D}, this selection approach is particularly promising. 
Upon searching the DESI AGN/QSO value-added catalog (Juneau et al. in preparation), we identified approximately 5,000 sources characterized by $M_\star<10^{10}~M_{\odot}$ and a [\iona{Ne}{v}] signal-to-noise ratio (S/N) greater than 10. Subsequent spectral fitting of this sample revealed roughly 140 sources exhibiting a [\iona{Ne}{v}] EW exceeding 30~\AA, although we emphasize that these results have not been closely examined to eliminate likely contamination due to, e.g., instrumental artifacts or inaccurate $M_\star$ measurements. This will allow our case study to expand into a population-level investigation, which we leave for future work. However, the detection rate for [\iona{Ne}{v}] remains remarkably low \citep{Reefe2022ApJ...936..140R,McKaig2024ApJ...976..130M}; for context, the SDSS DR8 catalog yields a detection rate of only 0.01\% among nearly one million galaxy spectra \citep{Reefe2022ApJ...936..140R}. This scarcity is likely a combined result of the intrinsic weakness of the [\iona{Ne}{v}] line, possible dust and gas attenuation on the host-galaxy scale, and the limitations of spectroscopic S/N in detecting [\iona{Ne}{v}].

\subsection{A Physical Scenario under the Merger-driven Framework}
\label{sec:mergepropp}
As shown in Figure~\ref{fig:image_decomposition}, the optical images of GAMA 376183 reveal a disturbed morphology. To quantify if it is consistent with mergers, we first calculate the classical morphological statistics such as Gini-$\rm M_{20}$ statistics \citep{Lotz2004AJ....128..163L}, concentration, asymmetry and smoothness statistics \citep{Bershady2000AJ....119.2645B,Conselice2003ApJS..147....1C,Lotz2004AJ....128..163L}, and outer asymmetry \citep{Wen2014ApJ...787..130W}, and they do not classify our source as a merger. However, these diagnostics are insensitive to late-stage mergers and minor companions, so they may not capture all merger signatures. Independently, the deep-learning merger classifier in \citet{Omori2023A&A...679A.142O} classifies our object as a merger with a precision of $\gtrsim80\%$. The following discussion will be carried out under the assumption that this galaxy is undergoing (or has undergone) a merger, and we acknowledge the caveat that the merger evidence solely based on the disturbed images may not be deterministic. Nevertheless, as will be shown below, the overall properties of our source are consistent with the expectations following mergers.\par
The merger-driven coevolution framework has been primarily proposed for supermassive black holes in massive galaxies \citep{Hopkins2006ApJS,Hopkins2008ApJS,Alexander2012NewAR}, where gas-rich mergers play a crucial role in triggering both intense star formation and rapid MBH growth. During the early post-merger stages, MBH accretion can even approach the Eddington limit and be heavily obscured by abundant fueling material. Eventually, radiation-driven outflows from the vicinity of the central MBH sweep out the obscuring matter, allowing the MBH to shine as an unobscured AGN.

Both numerical simulations and observations demonstrate the tendency for mergers to produce obscured AGNs. Hydrodynamic merger simulations predict sustained central inflows that increase the covering fraction of gas and dust, producing buried CT phases during peak accretion \citep[e.g.,][]{Hopkins2008ApJS,Blecha2018MNRAS.478.3056B,Kawaguchi2020ApJ...890..125K}. Observationally, post-mergers host a significantly higher fraction of mid-IR-selected AGNs than optical AGNs, suggesting that optically obscured AGNs become prevalent following mergers \citep{Koss2010ApJ...716L.125K,Satyapal2014MNRAS.441.1297S,Ellison2019MNRAS.487.2491E}. Furthermore, heavily obscured systems are more common in mergers than in isolated galaxies \citep{Kocevski2015ApJ...814..104K}. This merger-driven scenario is directly seen in extreme systems, such as high-$\lambda_\mathrm{Edd}$ dust-obscured galaxies \citep{Zou2020MNRAS.499.1823Z,Zou2025ApJ...990...94Z}, hot dust-obscured galaxies \citep{Wu2012ApJ,Assef2015ApJ...804...27A,Fan2016ApJ...822L..32F,Vito2018MNRAS}, and (ultra-)luminous infrared galaxies \citep{Sanders1988ApJ}.

Our morphological analyses of GAMA 376183 in Section~\ref{sec: imgdecomp} indicate that it is likely undergoing a merger. We verify that it also qualifies as a luminous infrared galaxy, which is observationally defined as those with IR luminosities above $10^{11}~L_{\odot}$ \citep{Sanders1996ARA&A..34..749S}, and Figure~\ref{fig:sed} shows that the luminous IR emission is significantly contributed by the AGN. Notably, up to $\sim$65\% of such luminous infrared galaxies in the local universe are associated with mergers \citep{Carpineti2015A&A...577A.119C}.

Further, our source is likely close to Eddington-limited. We estimate the black-hole mass $M_\mathrm{BH}$ using the $M_\mathrm{BH}{-}M_\star$ and $M_\mathrm{BH}{-}M_\mathrm{bulge}$ scaling relations from \citet{Kormendy2013ARA&A}, \citet{Reines2015ApJ}, and \citet{Greene20}, which provide consistent results. We adopt the bulge-based estimate of $\log M_\mathrm{BH}=6.83\pm0.30$, as the $M_\mathrm{BH}{-}M_\mathrm{bulge}$ relation exhibits a smaller intrinsic scatter.
Based on the SED-derived $L_\mathrm{bol}$, we calculate $\lambda_\mathrm{Edd}=0.81\pm0.33$. The implied X-ray bolometric correction factor, $\log k_\mathrm{bol}=\log L_\mathrm{bol}-\log L_\mathrm{X, int}=1.9$, agrees well with the expected value ($\log k_\mathrm{bol}=1.7$) from the $k_\mathrm{bol}-\lambda_\mathrm{Edd}$ relation in \citet{Duras2020A&A...636A..73D}, which has a scatter of 0.3~dex. 
All these parameters are listed in Table~\ref{tab:allprop}. Such heavily obscured, Eddington-limited accretion is a hallmark of post-merger stages \citep[e.g.,][]{Zou2020MNRAS.499.1823Z,Zou2025ApJ...990...94Z,Vito2018MNRAS}.

Regarding host star formation, mergers are expected to trigger starburst activity. While the delayed SFH model places the source on the star-forming main sequence, it fails to capture the enhanced UV emission. By contrast, adopting a SFH with a possible starburst provides a improved fit. This model not only better reproduces the observed SED, but also indicates that GAMA 376183 likely has experienced a burst of star formation, consistent with a merger-induced episode.

Overall, the properties of GAMA 376183 fit well within the merger-driven framework. This is notable, as it has been observationally unclear if this paradigm, originally developed for massive galaxies, applies to low-mass systems. Our source provides a compelling low-mass example demonstrating the physical link between mergers and heavily obscured, rapid accretion. At this stellar mass and redshift ($M_\star\sim10^{10}~M_\odot$, $z\sim0.2$), the merger rate for mass ratios above 1:10, the lower limit of the initial mass ratio of our source (see Section~\ref{sec: imgdecomp}), is about $0.05~\rm Gyr^{-1}$ \citep{Rodriguez2015MNRAS.449...49R}, indicating that such systems can exist in galaxies with $M_\star\sim10^{10}$ and are not ubiquitous. For comparison, the corresponding rate at $M_\star\sim10^{11}~M_\odot$ increases to $\sim0.1~\rm Gyr^{-1}$.  Intriguingly, the low-mass CT AGN candidate XMM-LSS 02399 from \citet{Zou2023ApJ}, discussed in Section~\ref{sec:Nevew}, is also heavily obscured, rapidly accretion, and likely a strong starburst galaxy. It is also likely hosted by a merging system (private communication). These two cases may represent a previously overlooked population of merger-driven AGNs in low-mass galaxies. Furthermore, the enhanced SFRs observed in our sample align with the findings of both \citet{Barchiesi2024A&A...685A.141B} and \citet{Vergani2018A&A...620A.193V}, who reported elevated SFRs in [\iona{Ne}{v}] emitters compared to the general population.

\section{Summary and Future Prospects}
\label{sec: summary}
In this work, we report the discovery of a powerful, heavily obscured AGN, GAMA 376183, hosted by a low-mass galaxy and potentially triggered by a merger, supported by NuSTAR observations and multiwavelength data. The main observational findings are as follows:

\begin{enumerate}
\item Optical spectroscopy reveals a remarkably strong [\iona{Ne}{v}] emission line (EW $=48$~\AA), indicative of heavy obscuration in X-rays. The large $N_\mathrm{H}$ derived from X-ray analysis supports the use of [\iona{Ne}{v}] as an effective tracer of AGN activity, even in cases of heavy obscuration and low host-galaxy masses (see Sections~\ref{sec: optspec} and \ref{sec:Nevew}).
\item High-resolution HSC imaging resolves multiple disturbed components, suggesting that GAMA 376183 is likely involved in an ongoing merger. Bulge--disk decomposition yields a bulge stellar mass of $M_{\rm bulge} = 10^{9.40}\ M_{\odot}$ and a disk stellar mass of $M_{\rm disk} = 10^{10.03}\ M_{\odot}$ (see Section~\ref{sec: imgdecomp}).

\item UV-to-IR SED fitting constrains the total stellar mass to $M_\star \approx 10^{10}\ M_{\odot}$ and the SFR to $8.36\ M_{\odot}\ \rm yr^{-1}$. The statistical improvement of the fit after adding a burst component to the SFH model supports the presence of a possible starburst episode. A high AGN fraction (0.89) and edge-on viewing angle ($89^{\circ}$) support its type 2 nature. Combining the SED-based bolometric luminosity and black-hole mass from scaling relations gives an Eddington ratio $\lambda_{\rm Edd} = 0.81$, indicative of rapid black-hole growth (see Sections~\ref{sec:SED} and \ref{sec:mergepropp}).
\item NuSTAR observations support that GAMA 376183 hosts a heavily obscured AGN ($\log N_{\rm H} = 23.3$). Its observed $2-10$~keV luminosity is significantly below the $L_{6\mu\mathrm{m}}$–$L_\mathrm{X}$ and $L_{\rm [Ne\; \text{\scshape v}]}$–$L_\mathrm{X}$ relations. The intrinsic $2$--$10$~keV luminosity ($L_{\rm X,int}=10^{42.9}~\rm erg~s^{-1}$) agrees with the expectation from $L_{6\mu\mathrm{m}}$ but is still somewhat lower than the $L_{\rm [Ne\; \text{\scshape v}]}$-based prediction, which may be due to enhanced \NeV emission in low-mass galaxies (see Sections~\ref{sec: xray} and \ref{sec: comp_lumin}).
\end{enumerate}

The combination of strong obscuration, rapid black-hole growth, a star-bursting host galaxy, and likely merger morphology suggests that GAMA 376183 is experiencing a rapid coevolution phase of its MBH and host galaxy.
Future systematic searches in large, deep spectroscopic surveys, such as DESI, will help build substantial samples of such systems, providing critical insight into AGN-galaxy coevolution among low-mass galaxies. 

\begin{acknowledgments}
We thank the referee for a thorough and constructive review. S.W. and X.W. are thankful for the support of the the National Key R\&D Program of China (No. 2025YFA1614100) and National Science Foundation of China (12133001). F.Z. is funded by NASA under award No. 80NSSC25K7085. W.N.B. thanks the Eberly Endowment at Penn State.
GAMA is a joint European-Australasian project based around a spectroscopic campaign using the Anglo-Australian Telescope. The GAMA input catalogue is based on data taken from the Sloan Digital Sky Survey and the UKIRT Infrared Deep Sky Survey. Complementary imaging of the GAMA regions is being obtained by a number of independent survey programmes, including GALEX MIS, VST KiDS, VISTA VIKING, WISE, Herschel-ATLAS, GMRT and ASKAP, providing UV to radio coverage. GAMA is funded by the STFC (UK), the ARC (Australia), the AAO, and the participating institutions. The GAMA website is \url{https://www.gama-survey.org/}.
\end{acknowledgments}

\software{Astropy \citep{Astropy2022ApJ...935..167A},
\texttt{QSOFITMORE} \citep{QSOFITMORE, QSOFIT2018ascl, Shen2019ApJS},
\texttt{GalfitS} (R. Li \& L. C. Ho, in preparation),
\texttt{CIGALE} \citep{CIGALE2019},
HEASoft \citep{HEAsoft2014ascl.soft08004N},
\texttt{LePhare} \citep{Arnouts99, Ilbert06},
\texttt{sherpa} \citep{Siemiginowska24},
\texttt{statmorph} \citep{statmor2019MNRAS.483.4140R}.
}
\bibliography{sample7}{}
\bibliographystyle{aasjournal}

\end{document}